\documentstyle[aps,twocolumn,epsfig]{revtex}
\begin{document}
\newcommand{\ve}[1]{\mbox{\boldmath $#1$}}
\twocolumn[\hsize\textwidth\columnwidth\hsize
\csname@twocolumnfalse%
\endcsname
 
\draft 
 
\title {Relaxation Processes in Clouds of Trapped Bosons above the 
Bose-Einstein Condensation Temperature}
\author{G. M. Kavoulakis$^1$, C. J. Pethick,$^{1,2}$ and H. Smith,$^3$}
\date{\today}
\address{$^1$Nordita, Blegdamsvej 17, DK-2100 Copenhagen \O, Denmark, \\
        $^2$Department of Physics, University of Illinois at
      Urbana-Champaign, 1110 West Green Street, Urbana, Illinois 
61801-3080, \\
        $^3$\O rsted Laboratory, H. C. \O rsted Institute,
         Universitetsparken 5, DK-2100 Copenhagen \O, Denmark}
\maketitle
 
\begin{abstract}

 We present a unified account of damping of low-lying collective modes
and of relaxation of temperature anisotropies in a trapped Bose gas in the
collisionless regime.  By means of variational techniques, we show that
the relaxation times for the two situations are closely related to the
simplest variational estimate of the viscous relaxation time.  We derive
rather precise theoretical expressions for the characteristic relaxation
times, and compare our results with experiment.

\end{abstract}
 
\vskip2pc]

   Many experiments have been reported recently on relaxation processes in
clouds of bosons in magnetic traps.  Among these are measurements of
damping of collective modes in vapors of rubidium \cite{jila1,jila2} and 
sodium \cite{mit,Ketnew}, and of relaxation of temperature anisotropies in 
cesium \cite{Monroe,dalibard}, rubidium \cite{Myatt,myatt}, and sodium 
vapors \cite{Davis}.  
It is an important task for theory to understand these processes 
for a number of 
reasons.  One is to interpret the experimental data, and to use them to
deduce atomic scattering properties.  The second is theoretical, since
these systems are qualitatively different from most other atomic systems
with which one is familiar, in that they are finite, and that the mean 
free path of an atom is, in the vast majority of experiments, comparable with
or larger than the dimensions of the cloud. One is thus not in the 
hydrodynamic regime, for which so much of the traditional kinetic theory 
of gases was developed. In attacking the problem of relaxation
in trapped clouds of atoms, one has to confront a number of fundamental
issues also of importance in other contexts, such as collective motion
in nuclei \cite{nuclei}. In this Letter we focus on the properties of
clouds above the Bose-Einstein condensation temperature, $T_c$. This work 
is of importance in its own right, as well as for understanding  physical 
processes below $T_c$.  

In the past, the theory of relaxation of temperature anisotropies in 
trapped clouds of alkali-atom vapors has been approached both analytically 
\cite{Myatt,myatt} and numerically, using Monte-Carlo simulations 
\cite{Monroe,dalibard} or other methods \cite{Wu},
and we know of no calculations of the damping 
of modes in other than the hydrodynamic limit \cite{KPS,Griffin}.  Here we 
shall attack the problems analytically, the basic physical idea being to 
consider the motion of atoms in the absence of collisions, and then to 
treat the effect of collisions as a perturbation.

We begin with some general considerations.
The traps employed in experiments performed to date have an axis of 
symmetry, which we shall refer to as the $z$ axis.  Generally the trapping
potential, $V(\bf r)$, is harmonic, and we shall write it in the form 
\begin{eqnarray}
  V= \frac 1 2 m[\omega_z^2 z^2+\omega_\perp^2 (x^2 +y^2)],
\label{potential}
\end{eqnarray}
where $\omega_z$ and $\omega_\perp$ are the frequencies associated with
motion of atoms in the trap along the axis of the trap, and perpendicular
to it.  The effect of the interaction energy on the mode frequency is
small, of relative order $n U_0/(k_B T)$, where $U_0=4\pi \hbar^2 a_{\rm 
sc}/m$ 
is the effective two-particle interaction. Here $n$ is the particle
density, $a_{\rm sc}$ is the scattering length for atom-atom
collisions, $m$ is the mass of an atom, and  $k_B$ is Boltzmann's constant.  
The quantity $n U_0/(k_B T)$ is
typically of order one per cent, and we shall neglect it.  In the absence
of collisions, motions along the symmetry axis and perpendicular to it are
independent, with modes associated with motion in the $z$ direction having
frequencies which are multiples of $\omega_z$, and those associated with
motion in the $x$ and $y$ directions having frequencies which are 
multiples of $\omega_{\perp}$.  Our approach is quite general, but since 
modes with frequency close to $2\omega_z$ and $2\omega_{\perp}$ have
been investigated in detail experimentally, we
shall focus on these.

Quantum mechanically, the spectrum of single-particle levels is discrete.  
However, the distribution of levels will be spread out due to deviations 
of the trap potential from the purely harmonic form, and due to 
interactions between particles.  We shall assume that for conditions of 
experimental relevance, with single-particle energies of order $10^2$ 
oscillator quanta, the spreading is sufficient that the particles may be 
described semiclassically, in terms of a distribution function $f(\bf r, 
\bf p)$, where $\bf p$ is the particle momentum.
We now calculate the damping of collective modes above $T_c$.
 
Since the system is dilute, in the sense that the typical particle 
separation is small compared with the atom-atom scattering length, the 
particle distribution function  
satisfies the Boltzmann equation, which is
\begin{eqnarray}
   \frac{\partial f}{\partial t} + \frac{\bf p}{m}\cdot {\ve \nabla} f  - 
 {\ve \nabla} V\cdot {\ve \nabla}_{\bf p} f =-I[f],
\end{eqnarray}
where $I[f]$ is the collision term.
Let us now consider an oscillation involving collective motion in the $z$ 
direction. In the absence of
collisions, the general solution of the Boltzmann equation having time 
dependence
$e^{-2i\omega_z t}$ is
\begin{eqnarray}
   \delta f = z_+^2 g(K_j),
\label{deltaf}
\end{eqnarray}
where  $z_\pm=p_z \pm i m \omega_z z$ are the classical equivalents
of the raising and lowering operators in quantum mechanics, $g$ is an 
arbitrary function, and
$K_j$ are constants of the motion.
Since $z_\pm$ varies in time as
$e^{\mp i\omega_z t}$, it is obvious that $\delta f$, 
Eq.\,(\ref{deltaf})
has the desired time dependence.  One may ask whether more
general forms are possible, based on, say $z_+^3z_-$.  However,
$z_+z_-= 2mE_z$ where
$E_i = p_i^2/(2m) + m\omega_i^2 x_i^2/2$ is the energy associated
with motion in the $i$ direction.   Thus this expression is still of the
general form given by Eq.\,(\ref{deltaf}).

In the absence of collisions there is enormous degeneracy among modes that
have frequency $2\omega_z.$  This corresponds in the quantum mechanics of
particles in a simple harmonic oscillator potential to the many different
ways of giving an excitation energy  $2\hbar \omega_z$ to the system.
This degeneracy is broken by collisions between particles, and we now wish
to find the mode which has the longest lifetime, because we anticipate
that this will be the one of experimental relevance.    Collisions will
also mix parts of the distribution with different time dependences in
the absence of collisions, {\it e.g.} those for which $\delta f \sim z_+^3
\sim e^{-3i\omega_z t}$.  However, since we are interested in the 
collisionless regime, the collision rate is small compared with the
oscillator frequency and this mixing will be small. It is
convenient to introduce the notation $\Phi({\bf r},{\bf p})=\delta f({\bf r},
{\bf p})/[f^0(1+f^0)]$, where $f^0$ is the equilibrium distribution 
function, in terms of which the Boltzmann equation is given by
\begin{eqnarray}
   \left( \frac {\partial}{\partial t} + i 2 \omega_z \right) 
 f^0(1+f^0)\Phi = -I[\Phi],
\end{eqnarray}
where the linearized collision integral is 
\begin{eqnarray}
   I[\Phi] = \int d\tau_1\int d\sigma|{\bf v}-{\bf v_1}|
  (\Phi+ \Phi_1- \Phi'- \Phi_1') \times \nonumber \\ 
 \times f^0 f^0_1 (1+{f^0}') (1+{f^0_1}').
\label{be}
\end{eqnarray}
Here the momenta of the incoming particles in a collision are denoted by $\bf 
p$ and $\bf p'$, and those of the outgoing ones by $\bf p_1$ and $\bf p'_1$, 
and $d\tau=d{\bf p}/(2\pi\hbar)^3$  denotes the 
volume element in momentum space. We have introduced
the differential cross section $d\sigma$, which in general depends
on the relative velocity $u=|{\bf v}-{\bf v_1}|$
of the two incoming particles as well as on
the angle between the relative velocity before and after
the collision. In this study we restrict ourselves to the case of $s$-wave
scattering, since the energies of the colliding particles are quite
low; we shall also assume for the moment that the cross section is 
independent of the momentum, which is a good approximation for  Rb and Na 
atoms, but later we shall
examine the case of a momentum-dependent cross section, which is relevant for 
Cs.

We now look for solutions of the Boltzmann equation having
the form $e^{-2i \omega_z t -\Gamma t}$.  These are just 
the eigenfunctions of the collision term, $\Psi_i$, which satisfy the 
equation 
\begin{eqnarray}
   \Gamma_i f^0(1+f^0)\Psi_i = I[\Psi_i],
\end{eqnarray}
where $\Gamma_i$ is the corresponding eigenvalue.  The most long-lived 
mode corresponds to the smallest value of $\Gamma_i$, which we denote by 
$\Gamma_0$. The task of finding $\Gamma_0$ is most simply attacked
by variational methods, since for any trial function $\Psi$ it
is clear that
\begin{eqnarray}
      \Gamma_0  \leq   \frac{\langle \Psi^*
    I[\Psi] \rangle}{\langle |\Psi|^2 f^0(1+f^0)  \rangle},
\label{assumption}
\end{eqnarray}
which is our basic variational expression for the damping rate.
Here the brackets denote integration over both coordinate and
momentum space. 

As a first application, we estimate the damping 
rate of the mode considered above.  We employ the simplest possible
trial function consistent with the fact that
it must be of the form of Eq.\,(\ref{deltaf}) for modes with frequency
$2 \omega_z$, and write $\Psi = z_+^2$. This trial function contains terms of 
the form $z^2$,
$zp_z$, and $p_z^2$. As is clear from Eq.\,(\ref{be}), the collision 
integral gives zero when operating on 
the first two of these because of conservation in collisions of particle 
number and momentum, and, because of energy conservation, it also gives zero 
when operating on $p^2$.  Consequently we may write 
\begin{eqnarray}
         \langle (z_+^*)^2I[z_+^2] \rangle =\langle p_z^2I[p_z^2] \rangle 
 =\langle (p_z^2-p^2/3)I[p_z^2-p^2/3] \rangle .       
\label{reduce}
\end{eqnarray}
This is a particularly convenient form because precisely the 
same quantity occurs in the simplest variational calculation of the 
shear viscosity, $\eta$, which is given by  \cite{Griffin,Henrik},   
\begin{eqnarray}
   \eta>\eta_{\rm var} = \frac 1 {m^2 k_B T} 
  \frac{[\int d\tau (p_z^2-p^2/3)^2 f^0(1+f^0)]^2}{\int d\tau 
(p_z^2-p^2/3)I[p_z^2-p^2/3]}.
\label{visc}
\end{eqnarray}  
Substitution of Eq.\,(\ref{reduce}) into Eq.\,(\ref{assumption}) leads to 
the result
\begin{eqnarray}
    \Gamma_0 \leq  \frac{ \langle
  (p_z^2-p^2/3)I[p_z^2-p^2/3] \rangle}
{\langle [p_z^2+(m\omega_z z)^2]^2 f^0(1+f^0)\rangle}.
\label{var}
\end{eqnarray}

Let us begin by considering a non-degenerate gas.  To bring out the 
physical content of our results we introduce the variational estimate for 
the viscous relaxation time.  This is given by $\tau_{\eta, {\rm 
var}}({\bf r}) =\eta_{\rm var}/[n({\bf r}) k_B T]$, where $n({\bf r})$ is 
the local density.  For the case of energy independent $s$-wave scattering 
one finds \cite{Henrik}
\begin{eqnarray}
    \tau_{{\eta},\rm{var}}^{-1} =  \frac {4 \sqrt 2} 5 n \sigma \bar{v},
\label{viscosity}
\end{eqnarray}
where $\sigma=8 \pi a_{\rm sc}^2$ is the total cross section, and 
$\bar{v}$ is the mean thermal velocity, $\bar{v} = (8 k_B T/\pi m)^{1/2}$.
From Eqs.\,(\ref{visc}) and (\ref{var}) we are led to the following 
expression for the damping 
\begin{eqnarray}
    \Gamma_0 \leq \frac{1}{6}\int d{\bf r}\,	\frac{n({\bf 
r})}{\tau_{{\eta},\rm{var}}} \left/ \int d{\bf r} \,n({\bf r}) \right..
\label{classical}
\end{eqnarray}
The factor of $1/6$ reflects the fact that collisions damp directly only 
the part of the distribution function varying as $p_z^2-p^2/3$, while all 
parts of the distribution contribute to the total energy associated with 
the oscillation. The average of the scattering rate that 
occurs here is proportional to $\int n^2({\bf r})
\, d{\bf r} / \int n({\bf r}) \, d{\bf r} =n(0)/2^{3/2}$, where $n(0)$ is 
the density at the center of the cloud, and therefore we find
\begin{eqnarray}
   \Gamma_0 \leq \frac 1 {12\sqrt 2} \, \frac 1 {\tau_{\eta,{\rm 
var}}(0)}
 = \frac 1 {15} n(0) \sigma \bar{v}.
\label{final}
\end{eqnarray}
One can systematically improve the estimates of mode damping by using more 
general trial functions.  As in the case of transport properties of classical 
gases, one finds that the simplest variational calculation is extremely 
good, and yields results which are within a few per cent of the exact answer.  
Details of these calculations will be reported elsewhere. 

Including effects of degeneracy to leading order
in $T_c/T$, we find
\begin{eqnarray}
     \Gamma_0 &\leq& \frac 1 {15} n_{\rm cl}(0) \sigma \bar{v} 
 \left[ 1 + \frac 3 {16} \zeta(3) \left( \frac {T_c} T \right)^3 \right].
\label{corr}
\end{eqnarray}
Here $n_{\rm cl}(0) = N  \omega_{\perp}^2 \omega_z [m /2 \pi k_B T]^{3/2}$, 
with $N$ being the total number of atoms, is the central 
density calculated neglecting effects of degeneracy, and $\zeta(\alpha)$ is 
the Riemann zeta 
function.
The coefficient $3 \zeta(3)/16$ is approximately equal to 0.225.

   We now consider oscillations in the $x$ and $y$ directions. An oscillation
in the $x$ direction alone is an eigenstate in the absence of collisions,
but when collisions are included, it will be strongly coupled to the 
degenerate oscillation in the $y$ direction. It is clear from the rotational
symmetry of the problem that the appropriate eigenstates are ones
having simple properties under rotation about the $z$ axis. The simplest
trial functions are thus $\Psi = x_+^2+y_+^2$
(corresponding to magnetic quantum number $m_z=0$) and $\Psi = x_+^2-y_+^2$
($m_z=2$), where $x_{\pm}$ and $y_{\pm}$ are defined by analogy with 
$z_{\pm}$.

    Starting with the $m_z=0$ mode, we see from Eq.\,(\ref{reduce}) 
that this choice for $\Psi$ gives the same answer for the collision 
integral as before,
since the only non-vanishing part comes from $p_x^2 + p_y^2$, which is
equal to $p^2 - p_z^2$. The denominator of Eq.\,(\ref{assumption}),
though, is twice as large in this case. Since the correction due
to the quantum degeneracy does not depend on the symmetry of the
mode, the damping rate is
\begin{eqnarray}
   \Gamma_0(m_z=0) &\leq& \frac 1 {30} n_{\rm cl}(0) \sigma \bar{v}
 \left[ 1 + \frac 3 {16} \zeta(3) \left( \frac {T_c} T \right)^3 \right].
\label{corrm0}
\end{eqnarray}
Concerning the $m_z=2$ mode, the part of $\Psi$ that
contributes to the collision integral is $p_x^2-p_y^2$, and since
$p_x^2-p_y^2 =(8 \pi/15)^{1/2} p^2 (Y_2^2+Y_2^{-2})$
and $p_x^2+p_y^2 -2p^2/3= (2/3)^{1/2}(8 \pi /15)^{1/2} p^2 Y_2^0$, where 
$Y_l^m$ are the
spherical harmonics, we find that $\Gamma_0(m_z=2) = 3 \, \Gamma_0(m_z=0)$, 
and thus,
\begin{eqnarray}
    \Gamma_0(m_z=2) &\leq& \frac 1 {10}  n_{\rm cl}(0) \sigma \bar{v}
  \left[ 1 + \frac 3 {16} \zeta(3) \left( \frac {T_c} T \right)^3 \right].
\label{corrm2}
\end{eqnarray}

   We now compare our results with experiment. The damping of oscillations 
along the axis of the trap has been measured by the MIT group 
\cite{mit}, and at $T=2 \, T_c$ they found a damping time of about $80$ ms. 
The
frequency, $\nu =\omega/2\pi$, of the mode is  $\approx 35$ Hz, somewhat below 
the frequency
$\nu_C = 2 \nu_z \approx 38$ Hz expected in the collisionless regime, but 
significantly above the value $\nu_H \approx \sqrt{12/5} \,\nu_z\approx 29$ Hz
in the hydrodynamic regime \cite{hy}. This is clear evidence that the system   
is in an intermediate regime between the collisionless and hydrodynamic 
limits,
but somewhat closer to the collisionless one.  
For $N = 5 \times 10^7$ at $T=2\,T_c$ \cite{private},
and for $\omega_{\perp} \approx 2 \pi \times 250$ Hz, $\omega_z \approx
2 \pi \times 19$ Hz,  
classically the central density is $\sim 2.9 \times 10^{13}$ cm$^{-3}$.
Taking  $a_{\rm sc} = 27.5$ \AA \cite{Dav}, we estimate that $\Gamma_0^{-1} 
\approx 48$ 
ms in the classical limit, or $\approx 46$ ms if the effects of degeneracy are 
included.  We can 
allow in an 
approximate fashion for the fact that the system is not in the collisionless 
limit by using the simple interpolation formula,
Eqs.\,(21) and (22) of Ref.\,\cite{KPS}, 
which indicates that the damping rate is approximately
$(\nu -\nu_H)/(\nu_C -\nu_H)$ times the value in the collisionless limit.  
This 
gives a theoretical estimate for the damping time of $\sim 70$ ms,  
in good agreement with the experimental value of $\sim 80$ ms. 

In a more recent experiment of the MIT group \cite{Ketnew},
axial oscillations 
were measured to have a damping time of $\approx 50$ ms 
at $T \approx T_c$ and a frequency  $\nu \approx 1.75 \, 
\nu_z$. In this experiment
$N \approx 8 \times 10^7$ at $T \approx T_c$, $\omega_{\perp}
\approx 2 \pi \times 230$ Hz, and $\omega_z \approx 2 \pi \times 17$ Hz.  Thus 
 $n_{\rm cl}(0) \approx 6.3 \times 10^{13}$ cm$^{-3}$ and $T_c 
\approx 1.9$ $\mu$K, so $\Gamma_0^{-1}(m=0) \approx 30$ ms for classical 
statistics.  The effects of degeneracy are more pronounced in this case, and 
if 
we take only the leading term, as in Eq.\,(\ref{corr}), we find a damping time 
of 
$\sim 24$ ms.  This experiment is slightly closer to the hydrodynamic limit 
than to the collisionless one, since the factor $(\nu -\nu_H)/(\nu_C -\nu_H)$ 
is  
$\approx 0.44$, and the damping time estimated as in the previous case
is 
$\sim 55$ ms, again in excellent agreement with the experimental
value, $\sim 50$ ms.

We now turn to measurements of modes involving motion perpendicular to the 
axis of the trap.   
The $m_z=0$ and  $m_z=2$ modes have both been investigated in experiments 
by the JILA group \cite{jila2}.  
 This experiment is in the collisionless regime, since the
product of the frequency $\omega$ of the mode
and the scattering time $\tau_{\rm sc}$ is $\approx 20$, and  
the mean free path is at least four times the typical dimension
of the cloud. Taking $N=8 \times 10^4$ at $T=1.1 \, T_c$,  
$\omega_{\perp} \approx 2 \pi \times 129$ Hz, and 
$\omega_z \approx 2 \pi \sqrt{8} \times 129$ Hz,  
we find that classically the central density is $\sim 3.7 \times 10^{13}$ 
cm$^{-3}$. Using a scattering length $a_{\rm sc} = 53$ \AA , we find from 
Eq.\,(\ref{corrm0}) that 
for the $m_z=0$ mode the damping time is $\Gamma_0^{-1}(m_z=0) \approx 100$ 
ms, while 
for the $m_z=2$ mode it is $\Gamma_0^{-1}(m_z=2) \approx 33$ ms.
Our results agree  in order of magnitude with the experimental results, but a 
more detailed comparison between theory and experiment should await a 
reappraisal of the experimental data.
   
 In experiments with cooled atoms, one often encounters situations where the
degrees of freedom associated with the motion 
of atoms in the various directions 
in the trap are out of equilibrium with each other 
\cite{Monroe,dalibard,Myatt,myatt,Davis}.
Crudely speaking, one may talk of anisotropies of the temperature of 
particles. We now show how temperature relaxation may 
be treated in our approach, and conclude that again the characteristic 
relaxation time is simply related to the  viscous relaxation time 
that determines the damping of oscillations.
If one changes the temperature associated with particle motion in the  
$z$ direction by an amount $\delta T$ and that associated with motion in 
the transverse direction by an amount $-\delta T/2$, thereby conserving the 
average particle energy, the change in the particle distribution function is 
proportional to $f^0(1+f^0)[E_z-(E_x+E_y)/2]\propto  z_+z_- - (x_+ x_- + y_+
y_-)/2$. The damping of such a disturbance may be treated in just the same 
way as for the case of oscillations, and one finds that the temperature 
relaxation
rate, $\Gamma_T$, is simply that given by Eq.\,(\ref{viscosity}), for a
density $n=n_{\rm cl}(0)/2^{3/2}$, multiplied by 1/2. The physical
reason for this factor of 1/2, is that both kinetic and 
potential energies have to be dissipated through collisions, whereas 
collisions degrade directly only the anisotropies associated with the kinetic 
energy. Therefore, we 
find for $\Gamma_T$ up to leading order in $T_c/T$
\begin{eqnarray}
  \Gamma_T \le  \frac 1 {5} n_{\rm cl}(0) \sigma \bar{v}
 \left[ 1 + \frac 3 {16} \zeta(3) \left( \frac {T_c} T \right)^3 \right].
\label{temprel3}
\end{eqnarray}
The classical version of the above equation can be extracted 
from Ref.\,\cite{myatt}. The number 5 which appears in the coefficient
of the above equation should be compared with the number 5.4
that results from a Monte Carlo analysis \cite{Monroe}.

   Our results may readily be generalized to more general forms of the 
scattering cross section, a situation relevant for Cs which has a two-body 
resonance close to zero energy \cite{dalibard}.  Assuming a total $s$-wave 
cross section $\sigma(u)$, one finds in the 
classical 
limit a temperature relaxation rate given by
\begin{eqnarray}
   \Gamma_T \leq \frac{1}{8 \sqrt{2}} n_{\rm cl}(0) \overline{\sigma(u) u},
\label{rate}
\end{eqnarray}
where  
\begin{equation}
  \overline{\sigma(u)u}  =\frac{\int_0^{\infty}du u^6\exp(-mu^2/4k_BT)  
\sigma(u) u}{\int_0^{\infty}du
 u^6\exp(-mu^2/4k_BT)}.
\label{gen}
\end{equation}
Under the experimental conditions \cite{dalibard} it is a rather good 
approximation to assume that the scattering is resonant, with $\sigma = 8 
\pi/k^2$ where $k=m 
u/(2\hbar)$ is the relative momentum.  We then obtain the 
result
 \begin{eqnarray}
   \Gamma_T \leq
  \frac {128} {15} \, \frac {n_{\rm cl}(0) \hbar^2}
{m^2 \bar{v}}.
\label{temprel4}
\end{eqnarray}
The coefficient 15 should be compared with the value
$2\times 10.7 \approx 21$ obtained from a Monte Carlo simulation
\cite{dalibard}.

    Helpful discussions with G. Baym, W. Ketterle, and C. E. Wieman
are gratefully acknowledged. We should also like to thank C. E. Wieman
for providing us with a copy of Ref.\,\cite{myatt}. 
G.M.K. would like to thank the Foundation
of Research and Technology, Hellas (FORTH) for its hospitality.

\end{document}